\documentclass[5p,times]{elsarticle}
\usepackage[dvips]{color}
\newcommand{\bea}{\begin{eqnarray}}
\newcommand{\eea}{\end{eqnarray}}
\newcommand{\be}{\begin{equation}}
\newcommand{\ee}{\end{equation}}
\newcommand{\np}{{\bf p}}
\newcommand{\hp}{\widehat{\bf p}}

\newcommand{\nh}{{\bf h}}

\newcommand{\nq}{{\bf q}}

\newcommand{\Qbar}{\not{\!Q}}

\newcommand{\kbar}{\not{\!k}}
\newcommand{\Pbar}{\not{\!P}}
\newcommand{\tauvec}{\mbox{\boldmath $\tau$}}

\newcommand{\Ivec}{\mbox{\boldmath $I$}}

\begin{document}

\title{
The frozen nucleon approximation in two-particle 
two-hole response functions
}

\author{I. Ruiz Simo}
\ead{ruizsig@ugr.es}
\author{J.E. Amaro}
\ead{amaro@ugr.es}
\address{Departamento de F\'{\i}sica At\'omica, Molecular y Nuclear,
and Instituto de F\'{\i}sica Te\'orica y Computacional Carlos I,
Universidad de Granada, Granada 18071, Spain}

\author{M.B. Barbaro}
\address{Dipartimento di Fisica, Universit\`a di Torino and
  INFN, Sezione di Torino, Via P. Giuria 1, 10125 Torino, Italy}

\author{J.A. Caballero}
\author{G.D. Megias}
  \address{Departamento de F\'{\i}sica At\'omica, Molecular y Nuclear,
Universidad de Sevilla, Apdo.1065, 41080 Sevilla, Spain}

\author{T.W. Donnelly}
\address{Center for Theoretical Physics, Laboratory for Nuclear
  Science and Department of Physics, Massachusetts Institute of Technology,
  Cambridge, MA 02139, USA}


\date{\today}


\begin{abstract}

We present a fast and efficient method to compute the
inclusive two-particle two-hole (2p-2h) electroweak responses 
in the neutrino and electron quasielastic inclusive cross
sections. The method is based on two approximations.  The first
neglects the  motion of the two initial nucleons
below the Fermi momentum, which are considered to be at rest. This
approximation, which is reasonable for high values of the momentum
transfer, turns out also to be quite good for moderate values of the
momentum transfer $q\gtrsim k_F$. The second approximation involves using
in the ``frozen'' meson-exchange currents (MEC) an effective 
 $\Delta$-propagator
averaged over the Fermi sea.  Within the resulting
``frozen nucleon approximation'', the inclusive 2p-2h responses are
accurately calculated with only a one-dimensional integral over the
emission angle of one of the final nucleons, thus drastically 
simplifying the calculation and reducing
the computational time. The latter makes this method especially
well-suited for implementation in Monte Carlo neutrino event
generators.
\end{abstract}


\maketitle

\section{Introduction}

The analysis of modern accelerator-based neutrino oscillation
experiments requires good control over the intermediate-energy
neutrino-nucleus scattering cross section \cite{Mos16,Kat17}.  In particular
the importance of multi-nucleon events has been suggested in many
calculations of charge-changing quasielastic cross sections
$(\nu_\mu,\mu)$, at typical neutrino energies of $\sim 1$
GeV~\cite{Nie11,Nie12,Gra13,Ama11,Ama12,Mar09,Mar10}.  The
contribution of two-particle-two-hole (2p-2h) excitations is now thought
to be essential for a proper description of
data~\cite{Agu10,Nak11,And12,Abe13,Fio13,Abe14,Wal15,Ank16,Rod16}.
Thus a growing interest has arisen in including 2p-2h models into the
Monte Carlo event generators used by the neutrino
collaborations
\cite{Hayato:2009zz,Andreopoulos:2009zz,Andreopoulos:2009rq,Sobczyk:2008zz}.

The only 2p-2h model implemented up to date in some of the Monte Carlo
neutrino event generators corresponds to the so-called 'IFIC Valencia
model'~\cite{Nieves:2011pp,Gran:2013kda}, which has been incorporated
in GENIE~\cite{Schwehr:2016pvn}. There are also plans to incorporate
the 'Lyon model'~\cite{Martini:2009uj} in GENIE, while
phenomenological approaches like the effective transverse enhancement
model of~\cite{Bodek:2011ps} are implemented, for instance, in
NuWro generator~\cite{Zmuda:2015twa}.
 
 One of the main problems to implementing the 2p-2h models is the high
 computational time. This is due to the large number of nested
 integrals involved in the evaluation of the inclusive hadronic tensor
 with sums over the final 2p-2h states. To speed up the calculations,
 several approximations can be made, such as choosing an average momentum
 for the nucleons in the local Fermi gas~\cite{Nieves:2011pp},
 neglecting the exchange matrix elements, or reducing the number of
 integrations to two nested integrals by performing a non-relativistic
 expansion of the current operators~\cite{VanOrden:1980tg}.  The
 latter approach is only useful for some pieces of the elementary
 2p-2h response.

 In this work we present a fast and very efficient method to calculate
 the inclusive 2p-2h responses in the relativistic Fermi gas model
 (RFG). This approach, denoted as the frozen nucleon approximation, was
 first explored in \cite{Simo:2014wka} but restricted to the analysis
 of the 2p-2h phase-space. Here it is extended to the evaluation of
 the full hadronic tensor assuming that the initial momenta of the two
 struck  nucleons can be neglected for high enough energy and momentum
 transfer, $q>k_F$.  The frozen nucleon approximation was found to
 work properly in computing the phase space function for two-particle
 emission in the range of momentum transfers of interest for neutrino
 experiments with accelerators. Here we investigate the validity of
 the frozen approximation beyond the phase-space study by including the
 electroweak meson-exchange current (MEC) model of
 \cite{Simo:2016ikv}.  We find that the presence of virtual delta
 excitations requires one to introduce a ``frozen'' $\Delta$-propagator,
 designed by a convenient average over the Fermi sea.
 
The main advantage of the frozen approximation consists in reducing
 the number of nested integrals needed to evaluate the inclusive 2p-2h
 electroweak responses from 7 (full calculation) to 1. Thus it is well-suited to computing
 the 2p-2h neutrino cross sections folded with the neutrino
 flux, and it can be of great help in order to implement the 2p-2h models in the
 Monte Carlo codes currently available.
 
 The plan of this work is as follows: in section \ref{sec_form} we
 review the formalism of neutrino scattering and describe
 mathematically the frozen approximation approach. In section
 \ref{sec_results} we validate the nucleon frozen approximation by
 computing the 2p-2h response functions and by comparing with the
 exact calculation. Finally, in section \ref{sec_conclusions} we
 summarize our conclusions.
 
\section{Formalism}
\label{sec_form}

\subsection{Cross section and hadronic tensor}

The double-differential inclusive $(\nu_l,l^-)$ or
$(\bar{\nu}_l, l^+)$ cross section is given by
\begin{eqnarray} \label{cross}
\frac{d^2\sigma}{d\Omega'd\epsilon'}
&=&
 \sigma_0
\left[
\widetilde{V}_{CC} R^{CC}
+ 2 \widetilde{V}_{CL} R^{CL}
+ \widetilde{V}_{LL} R^{LL}
\right.
\nonumber\\
&&
\left.
+ \widetilde{V}_{T} R^{T}
\pm 2 \widetilde{V}_{T'} R^{T'}
\right] \, ,
\end{eqnarray}
where the sign $\pm$ is positive for neutrinos and negative for
antineutrinos.  The term $\sigma_0$ in Eq.~(\ref{cross}) represents the
elementary neutrino scattering cross section with a point nucleon,
while the $\widetilde{V}_K$ 
are kinematic factors that 
depend on lepton kinematic variables. Their explicit expressions
can be found in \cite{Amaro:2005dn}. The relevant nuclear
physics is contained in the five nuclear response functions
$R^K(q,\omega)$, where $\nq$ is the momentum transfer, defining the
$z$ direction, and $\omega$ is the energy transfer.  They are defined as
suitable combinations of the hadronic tensor
\begin{eqnarray}
R^{CC} &=& R^L = W^{00} \label{rcc} \\
R^{CL} &=& -\frac12\left(W^{03}+ W^{30}\right) \\
R^{LL} &=& W^{33}  \\
R^{T} &=& W^{11}+ W^{22} \\
R^{T'} &=& -\frac{i}{2}\left(W^{12}- W^{21}\right)\,. \label{rtprima}
\end{eqnarray}

In this work we compute the inclusive hadronic tensor for two-nucleon
emission in the relativistic Fermi gas, given by
 \begin{eqnarray}
&&W^{\mu\nu}_{\rm2p-2h}
=
\frac{V}{(2\pi)^9}\int
d^3p'_1
d^3h_1
d^3h_2
\frac{m_N^4}{E_1E_2E'_1E'_2}
\nonumber \\ 
&&
\times r^{\mu\nu}(\np'_1,\np'_2,\nh_1,\nh_2)\;
\delta(E'_1+E'_2-E_1-E_2-\omega)
\nonumber\\
&&
\times \Theta(p'_1,p'_2,h_1,h_2)\, , \;
\label{hadronic}
\end{eqnarray}
where $\bf p'_2= h_1+h_2+q-p'_1$ by momentum conservation, $m_N$ is
the nucleon mass, $V$ is the volume of the system and we have defined
the product of step functions
\begin{eqnarray}
\kern -8mm
\Theta(p'_1,p'_2,h_1,h_2) &=&
\theta(p'_2-k_F)
\theta(p'_1-k_F)
\theta(k_F-h_1)
\theta(k_F-h_2) \nonumber \\
&&
\end{eqnarray}
with $k_F$ the Fermi momentum. 

Finally the function $r^{\mu\nu}(\np'_1,\np'_2,\nh_1,\nh_2)$ is the
elementary hadron tensor for the 2p-2h transition of a nucleon pair
with given initial and final momenta, summed up over spin and
isospin,
\begin{eqnarray}
\kern -8mm
r^{\mu\nu}(\np'_1,\np'_2,\nh_1,\nh_2) &=& \frac{1}{4}\sum_{s,t}
j^{\mu}(1',2',1,2)^*_A
j^{\nu}(1',2',1,2)_A \, , \nonumber \\
&&
\label{elementary}
\end{eqnarray}
which is written in terms of the antisymmetrized two-body current
matrix elements 
\begin{equation} \label{anti}
j^{\mu}(1',2',1,2)_A
\equiv j^{\mu}(1',2',1,2)-
j^{\mu}(1',2',2,1) \,.
\end{equation}
 The factor $1/4$ in Eq.~(\ref{elementary}) accounts for the
 antisymmetry of the
 two-body wave function.  

For the inclusive responses
 considered in this work there is a global axial symmetry, so we can
 fix the azimuthal angle of one of the particles.  We choose
 $\phi'_1=0$, and consequently the integral over $\phi'_1$ gives a factor
 $2\pi$. Furthermore, the energy delta function enables
analytical integration over $p'_1$, and so the integral in
Eq.~(\ref{hadronic}) can be reduced to 7 dimensions (7D). In the ``exact''
results shown in the next section, this 7D integral has been computed
numerically using the method described in \cite{Simo:2014wka}.

\subsection{Frozen nucleon approximation}

The frozen nucleon approximation consists in assuming that 
the momenta of the initial nucleons can be neglected
for high enough  values of the momentum transfer. 
Thus, in the integrand of Eq.~(\ref{hadronic}), we set
$\nh_1=\nh_2=0$, and $E_1=E_2=m_N$.
We roughly expect this
 approximation
to become more accurate as the momentum transfer increases.
 The integration over $\nh_1,\nh_2$ is trivially performed and 
the response function $R^K$, with $K=CC, CL, LL, T, T'$,
is hence approximated by
\begin{eqnarray}
\kern -4mm 
R^{K}_{\rm frozen}
&=& \frac{V}{(2\pi)^9}
\left(\frac43\pi k_F^3\right)^2
\int d^3p'_1\;
\frac{m_N^2}{E'_1E'_2}\;
r^{K}(\np'_1,\np'_2,\mathbf{0},\mathbf{0})
\nonumber\\
&\times&
\delta(E'_1+E'_2-2m_N-\omega)\;
\Theta(p'_1,p'_2,0,0) \,,
\label{tensor_frozen}
\end{eqnarray}
where $\np^\prime_2=\nq-\np^\prime_1$ and $r^{K}$ are the elementary
response functions for a nucleon pair excitation, which are defined
similarly to Eqs. (\ref{rcc}--\ref{rtprima}).  The integral over
$p^\prime_1$ can be done analytically by using the delta function for
energy conservation, and the integral over $\phi^\prime_1$ gives again
a factor of $2\pi$.  Thus only an integral over the polar angle
$\theta^\prime_1$ remains:
\begin{eqnarray}
&& R^{K}_{\rm frozen} =
\frac{V}{(2\pi)^9}\,
\left(\frac43\pi k_F^3\right)^2
2\pi 
\int_0^\pi
d\theta'_1\sin\theta'_1
\nonumber
\\
&\times &
\sum_{\alpha=\pm}
\left.
\frac{m_N^2 p'_1{}^2\Theta(p'_1,p'_2,0,0)}
{E'_1E'_2 \left| \frac{p'_1}{E'_1}-\frac{\np'_2\cdot\hp'_1}{E'_2} \right|}\,
r^{K}(\np'_1,\np'_2,\mathbf{0},\mathbf{0})
\right|_{p'_1= p'_1{}^{(\alpha)}} 
\label{frozen_eq}
\end{eqnarray}
where the sum runs over the, in general two, possible 
values of the momentum of the first particle for
given emission angle $\theta'_1$. These are obtained as the positive solutions
$p'_1{}^{(\pm)}$  of the energy conservation equation
\begin{equation} 
2m_N+\omega= \sqrt{p'_1{}^2+m_N^2}+\sqrt{(\nq-\np'_1)^2+m_N^2}\,.
\end{equation}
The explicit values of the solutions of the above equation can be
found in the appendix of~\cite{Simo:2014wka}.
Care is needed in performing the integral over $\theta'_1$ because the
denominator inside the integral can be zero for some kinematics. The
quadrature in these cases can be done with the methods explained in
\cite{Simo:2014wka,Simo:2014esa}.

\begin{figure}
\centering
\includegraphics[width=8cm,bb=160 190 455 690]{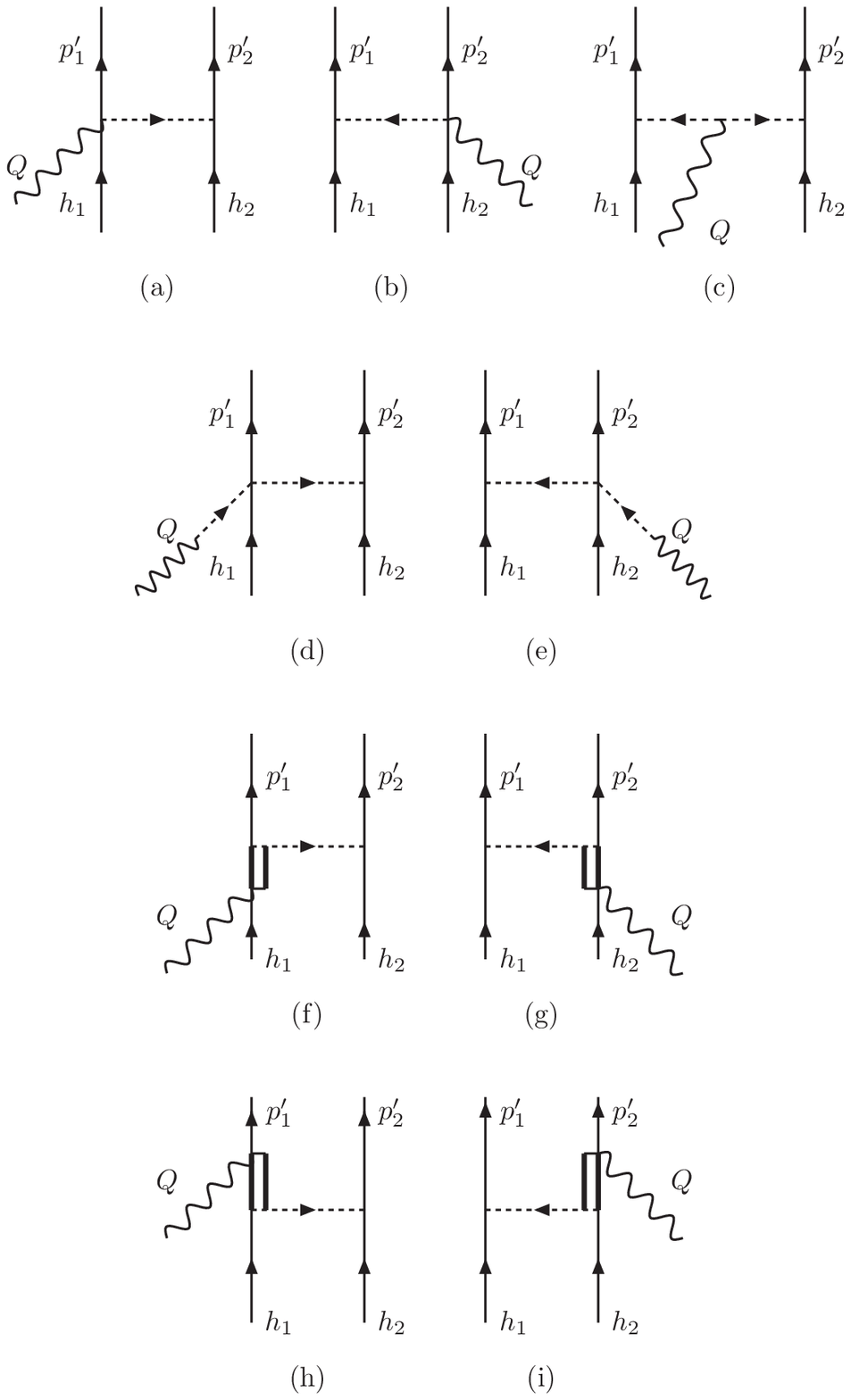}
\caption{Feynman diagrams for the electroweak MEC model used in 
this work.}\label{fig_feynman}
\end{figure}

\subsection{Electroweak meson-exchange currents}

To investigate the validity of the frozen nucleon approximation, we
have to choose a specific model for the two-body current matrix elements
$j^{\mu}(1',2',1,2)$ 
entering in the elementary 2p-2h response functions, 
Eqs.~(\ref{elementary},\ref{anti}). 
Here  we use the relativistic
model of electroweak MEC operators developed in \cite{Simo:2016ikv}. 
The MEC model can be summarized by the Feynman diagrams depicted in Fig.~\ref{fig_feynman}. 
It comprises several contributions coming from 
the pion production amplitudes of \cite{Hernandez:2007qq}.

The Seagull current, corresponding to diagrams (a,b), 
 is given by the sum of vector and axial-vector pieces 
 \begin{eqnarray}
 j^\mu_{\rm sea}&=&
  \left[I_V^{\pm}\right]_{1'2',12}
\frac{f^2_{\pi NN}}{m^2_\pi}
\frac{\bar{u}_{s^\prime_1}(\np^\prime_1)\,\gamma_5
 \kbar_{1} \, u_{s_1}(\nh_1)}{k^2_{1}-m^2_\pi}\,
 \nonumber\\
 &\times& 
 \bar{u}_{s^\prime_2}(\np^\prime_2)
 \left[ F^V_1(Q^2)\gamma_5 \gamma^\mu
 + \frac{F_\rho\left(k_{2}^2\right)}{g_A}\,\gamma^\mu
   \right] u_{s_2}(\nh_2)
   \nonumber\\
   &+&
   (1\leftrightarrow2) \,,
\label{seacur}
 \end{eqnarray}
where  $I_V^{\pm}=(I_V)_x\pm i (I_V)_y$
corresponds to the $\pm$-components of the two-body isovector operator
$\Ivec_V = i \left[\tauvec(1) \times\tauvec(2)\right]$.
 The $+ (-)$ sign refers to 
 neutrino (antineutrino) scattering.
 The four-vector $k^\mu_1=(p^\prime_1-h_1)^\mu$ 
is the momentum  carried by the exchanged pion
and $Q^\mu=(\omega,\nq)$.
 The $\pi NN$ ($f_{\pi NN}=1$) and axial ($g_A=1.26$) couplings,
and the form factors ($F_1^V$, $F_{\rho}$) have been taken from the 
pion production amplitudes of \cite{Hernandez:2007qq}.

The  Pion-in-flight current corresponding to diagram (c) 
is purely vector and is given by
  \begin{eqnarray}
 j^\mu_{\pi}&=& \left[I_V^{\pm}\right]_{1'2',12}
 \frac{f^2_{\pi NN}}{m^2_\pi}\,
 \frac{ F^V_1(Q^2) \,
\left(k^\mu_{1}-k^\mu_{2}\right)}{\left(k^2_{1}-m^2_\pi\right)
 \left(k^2_{2}-m^2_\pi\right)}\, \nonumber
 \\
&\times& 
\bar{u}_{s^\prime_1}(\np^\prime_1)\,\gamma_5
 \kbar_{1} \, u_{s_1}(\nh_1)
\bar{u}_{s^\prime_2}(\np^\prime_2)\,\gamma_5
 \kbar_{2} \, u_{s_2}(\nh_2) \, ,
\end{eqnarray}
where $k_2^\mu = (p'_2-h_2)^\mu $ is the momentum of the pion absorbed
by the second nucleon.

The pion-pole current corresponds to diagrams (d,e) and is purely axial,
given by
\begin{eqnarray}
j^\mu_{\rm pole}&=&\left[I_V^{\pm}\right]_{1'2',12}
\frac{f^2_{\pi NN}}{m^2_\pi}\,
\frac{F_\rho\left(k_{1}^2\right)}{g_A}\;Q^\mu
\bar{u}_{s^\prime_1}(\np^\prime_1)\Qbar u_{s_1}(\nh_1)
\nonumber\\
&\times&
\frac{
\bar{u}_{s^\prime_2}(\np^\prime_2)\,\gamma_5 \kbar_{2} \, u_{s_2}(\nh_2)
 }{\left(k^2_{2}-m^2_\pi\right)\left(Q^2-m^2_\pi\right)}
+(1\leftrightarrow2) \,.
\end{eqnarray}

Finally 
the $\Delta$ current corresponds
in Fig.~\ref{fig_feynman} to diagrams (f, g) for the
forward and (h, i) for the backward $\Delta$ propagations, respectively.
The current matrix elements are given by
\begin{eqnarray}
j^\mu_{\Delta}&=&j^\mu_{\Delta,\rm F}+
j^\mu_{\Delta,\rm B}
\\
j^\mu_{\Delta,\rm F}&=&
\frac{f^* f_{\pi NN}}{m^2_\pi}\,
\left[U_{\rm F}^{\pm}\right]_{1'2',12}
\frac{\bar{u}_{s^\prime_2}(\np^\prime_2)\,\gamma_5
 \kbar_{2} \, u_{s_2}(\nh_2)}{k^2_{2}-m^2_\pi}
\nonumber\\
&\times& k^\alpha_{2}\,\bar{u}_{s^\prime_1}(\np^\prime_1)G_{\alpha\beta}(h_1+Q)
\Gamma^{\beta\mu}(h_1,Q)u_{s_1}(\nh_1)
\nonumber\\
&+&(1\leftrightarrow2)
\label{delta_forward}
\\
j^\mu_{\Delta,\rm B}&=&
\frac{f^* f_{\pi NN}}{m^2_\pi}\,
\left[U_{\rm B}^{\pm}\right]_{1'2',12}\; 
\frac{\bar{u}_{s^\prime_2}(\np^\prime_2)\,\gamma_5
 \kbar_{2} \, u_{s_2}(\nh_2)}{k^2_{2}-m^2_\pi}
\nonumber\\
&\times& k^\beta_{2}\,\bar{u}_{s^\prime_1}(\np^\prime_1)
\hat{\Gamma}^{\mu\alpha}(p^\prime_1,Q)
G_{\alpha\beta}(p^\prime_1-Q)u_{s_1}(\nh_1)
\nonumber\\
&+&(1\leftrightarrow2)\label{delta_backward} \, .
\end{eqnarray}
The $\pi N\Delta$ coupling is $f^*=2.13$. 
The forward, 
$U_{\rm F}^{\pm}= U_{Fx}\pm i U_{Fy}$, and
backward, 
$U_{\rm B}^{\pm}= U_{Bx}\pm i U_{By}$, 
isospin transition operators have the following cartesian components
\begin{eqnarray}
U_{\rm Fj}&=&\sqrt{\frac32}
\sum_i\left(T_i T_j^\dagger\right)\otimes
\tau_i\\
U_{\rm Bj}&=&
\sqrt{\frac32}
\sum_i\left(T_{j}\, T^\dagger_i\right)\otimes
\tau_i,
\end{eqnarray}
where $\vec{T}$ and $\vec{T}^\dagger$ are the isovector transition
operators from isospin $\frac32$ to $\frac12$ or vice-versa,
respectively.  The  $+(-)$ operator 
is for neutrino (antineutrino) scattering.

The $\Delta$-propagator, $G_{\alpha\beta}(P)$, is given by
\begin{equation}\label{delta_prop}
 G_{\alpha\beta}(P)= \frac{{\cal P}_{\alpha\beta}(P)}{P^2-
 M^2_\Delta+i M_\Delta \Gamma_\Delta+
 \frac{\Gamma^2_{\Delta}}{4}} \, ,
\end{equation}
where ${\cal P}_{\alpha\beta}(P)$ is  the projector over
spin-$\frac32$ on-shell particles,
\begin{eqnarray}
{\cal P}_{\alpha\beta}(P)&=&-(\Pbar+M_\Delta)
\left[g_{\alpha\beta}-\frac13\gamma_\alpha\gamma_\beta-
\frac23\frac{P_\alpha P_\beta}{M^2_\Delta}\right.
\nonumber\\
&+&\left.
\frac13\frac{P_\alpha\gamma_\beta-
P_\beta\gamma_\alpha}{M_\Delta}\right]
\end{eqnarray}
and whose denominator has been obtained from the free propagator for
stable particles, $\frac{1}{P^2-M^2_\Delta}$, with the replacement
$M_\Delta \rightarrow M_\Delta -i \frac{\Gamma_\Delta}{2}$ to take
into account the finite decay width of the $\Delta\,(1232)$.

The tensor $\Gamma^{\beta\mu}(P, Q)$ 
in the forward current is the weak
$N\rightarrow\Delta$ transition vertex 
---a combination of gamma matrices  with vector
and axial-vector contributions:
\begin{eqnarray}
\Gamma^{\beta\mu}(P,Q)&=&\Gamma^{\beta\mu}_V(P,Q)
+\Gamma^{\beta\mu}_A(P,Q)
\\
\Gamma^{\beta\mu}_V(P,Q)&=&\left[\frac{C^V_3}{m_N}
\left(g^{\beta\mu}\Qbar-Q^\beta\gamma^\mu\right)\right.
\nonumber\\
&+&\left.\frac{C^V_4}{m^2_N}\left(g^{\beta\mu}Q\cdot \left(P+Q\right)-
Q^\beta \left(P+Q\right)^\mu\right)\right.
\nonumber\\
&+&\left.\frac{C^V_5}{m^2_N}\left(g^{\beta\mu}
Q\cdot P-Q^\beta P^\mu\right)+C^V_6 
g^{\beta\mu}\right]\gamma_5
\\
\Gamma^{\beta\mu}_A(P,Q)&=&
\frac{C^A_3}{m_N}\left(g^{\beta\mu}\Qbar-
Q^\beta\gamma^\mu\right)\nonumber\\
&+& \frac{C^A_4}{m^2_N}\left(g^{\beta\mu}Q\cdot \left(P+Q\right)-
Q^\beta \left(P+Q\right)^\mu\right) \nonumber\\
&+& C^A_5 g^{\beta\mu}+\frac{C^A_6}{m^2_N} 
Q^\beta Q^\mu \, .
\end{eqnarray}
For the backward current, we take
\begin{equation}
\hat{\Gamma}^{\mu\alpha}(P^\prime, Q)=\gamma^0
\left[\Gamma^{\alpha\mu}(P^\prime,-Q)\right]^{\dagger}
\gamma^0 \, .
\end{equation}
Finally, it is worth noting that the form factors $C^{V,A}_i$ 
are taken from \cite{Hernandez:2007qq}.
We refer to that work for further details of the model.

\subsection{The frozen $\Delta$-propagator}

The evaluation of the relevant elementary responses requires one to
contract the electroweak two-body MEC with themselves by spin-isospin
summation. This leads to the squares of each of the diagrams depicted
in Fig.~\ref{fig_feynman} plus all their interferences.

The validity of the frozen nucleon approximation relies on the fact
that the integrand inside the 2p-2h response is a function that
depends slowly on the momenta of the two initial nucleons inside the
Fermi sea. In that case the mean-value theorem applied to the
resolution of the integrals provides very precise results.
This is so for all of the diagrams
of the MEC except for the forward $\Delta$ diagram, which shows a sharp
maximum for kinematics around the $\Delta$ peak for pion emission,
located at $\omega=\sqrt{q^2+m_\Delta^2}-m_N$. This is due to the denominator 
in the $\Delta$ propagator,
\begin{equation}
G_{\Delta}(H+Q)
\equiv
\frac{1}{(H+Q)^2-
 M^2_\Delta+i M_\Delta \Gamma_\Delta+
 \frac{\Gamma^2_{\Delta}}{4}} \, ,
\label{denominator}
\end{equation}
where $H^\mu=(E_{\nh},\nh)$ is the momentum of the hole that gets
excited to a $\Delta$.  

In these cases the integrand changes very significantly with a small
variation of the momentum of the holes and consequently, the frozen
approximation cannot properly describe the integrand. On the contrary,
it only provides a general estimation of the order of magnitude.  To
get rid of these difficulties we have developed a prescription to deal
with the forward $\Delta$-propagator appearing in
Eq.~(\ref{delta_forward}).  This procedure is based on the
use of an effective propagator (``frozen'' ) for the
$\Delta$, conveniently averaged over the Fermi gas. This average is 
an analytical complex function, which is used instead of the ``bare''
propagator inside the frozen approximation, recovering the precision
of the rest of diagrams.

The ``frozen'' prescription amounts to the replacement:
\begin{equation}
G_{\Delta}(H+Q) \rightarrow G_{\rm frozen}(Q) \, ,
\end{equation}
where the frozen denominator is defined by
\begin{equation}
G_{\rm frozen}(Q)
=
\frac{ \int d^3h  \theta(k_F-\left|\nh\right|)
G_{\Delta}(H+Q)
}{\frac43 \pi k^3_F} \,.
\label{replacement_rule}
\end{equation}

Taking the non-relativistic
limit for the energies of the holes ($E_{\nh}\simeq m_N$), 
which is justified because hole momenta are below the Fermi
momentum, itself a value far below the nucleon rest mass, we 
can write:
\begin{equation}\label{averaged_prop}
G_{\rm frozen}(Q)
=
\frac{1}{\frac43 \pi k^3_F}
\int 
 \frac{d^3h \;\theta(k_F-\left|\nh\right|)}{a-2\,\nh\cdot\nq+ib} \, ,
\end{equation}
where
\begin{eqnarray}
a&\equiv& m^2_N+Q^2+2m_N\omega-M^2_\Delta+
\frac{\Gamma^2_\Delta}{4}\label{a}\\
b&\equiv& M_\Delta \Gamma_\Delta\label{b} \,.
\end{eqnarray}

Assuming the $\Delta$ width
($\Gamma_\Delta$) to be constant, we can 
integrate Eq.~(\ref{averaged_prop}) over the angles, getting
\begin{equation}
G_{\rm frozen}(Q)
= 
\frac{1}{\frac43 \pi k^3_F} 
\frac{\pi}{q}
\int^{k_F}_{0} dh h 
\ln\left[
\frac{a+2hq+ib}{a-2hq+ib}
\right] \,.
\end{equation}
Note the complex logarithm inside the integral, 
which provides the needed kinematical 
dependence of the averaged propagator,
differing from the bare Lorentzian shape. 
Finally the integral over the momentum $h$ can also be performed, resulting in 
\begin{eqnarray}
\kern -5mm  G_{\rm frozen}(Q)
&=& 
\frac{1}{\frac43 \pi k^3_F} 
\frac{\pi}{q}\left\{ \frac{\left(a+ib\right)k_F}{2q}
\right.
\label{averaged_denom}\\
&&
\kern -1cm \mbox{}+\left.
\frac{4q^2k^2_F - (a+ib)^2}{8q^2}
\ln\left[\frac{a+2k_Fq+ib}{a-2k_Fq+ib}\right]
\right\} \, .
\nonumber
\end{eqnarray}

By comparing the response functions evaluated in the frozen
approximation, {\it i.e.,} substituting the denominator of the $\Delta$
propagator in Eq.~(\ref{delta_prop}) for the frozen expression in 
Eq.~(\ref{averaged_denom}), with the exact results, we find that the
shapes around the $\Delta$ peak are similar, but with slightly
different width and position of the center of the peak.  We have
checked that the differences can be minimized by changing the
parameters $a,b$ with respect to the ``bare'' ones, given by
Eqs.~(\ref{a},\ref{b}). This is because we have computed the averaged
denominator without taking into account the current matrix elements
appearing in the exact responses, although the functional form and
kinematical dependence is the appropriate one.

In practice, we adjust $\Gamma_\Delta$ and apply a shift
in the expression for $a$ in Eq.~(\ref{a}) in order to
obtain the best approximation to the exact results. 
The effective ``frozen'' parameters we actually introduce in
Eq.~(\ref{averaged_denom}), are  given by
\begin{eqnarray}
\kern -0.8cm 
a_{\rm frozen}&\equiv& m^2_N+Q^2+2m_N(\omega+\Sigma_{\rm frozen})-M^2_\Delta+
\frac{\Gamma^2_{\rm frozen}}{4}
\label{afrozen}\\
\kern -0.8cm 
b_{\rm frozen}&\equiv& M_\Delta \Gamma_{\rm frozen}
\label{bfrozen} \,.
\end{eqnarray}
We consider  $\Gamma_{\rm frozen}$ and the frozen shift, 
$\Sigma_{\rm frozen}$, to be tunable 
parameters depending on the momentum transfer $q$.
We have adjusted these parameters 
 for different 
$q$-values and we provide them in Table~\ref{table1}.

\begin{table}[h]
\centering
\begin{tabular}{|c|c|c|}\hline 
$q$ (MeV/c) & $\Sigma_{\rm frozen}$ (MeV) & $\Gamma_{\rm frozen}$ (MeV) \\\hline 
300 & 20 & 130 \\\hline
400 & 65 & 147 \\\hline
500 & 65 & 145 \\\hline 
800 & 80 & 125 \\\hline 
1000 & 100 & 100 \\\hline 
1200 & 115 & 85 \\\hline 
1500 & 150 & 40 \\\hline 
2000 & 150 & 0   \\\hline
\end{tabular}
\caption{Values of the free parameters of the Fermi-averaged
$\Delta$-propagator for different kinematic situations corresponding
to different values of the momentum transfer $q$.}\label{table1}
\end{table}

\begin{figure}[h]
\centering
\includegraphics[width=8cm,bb=125 535 480 780]{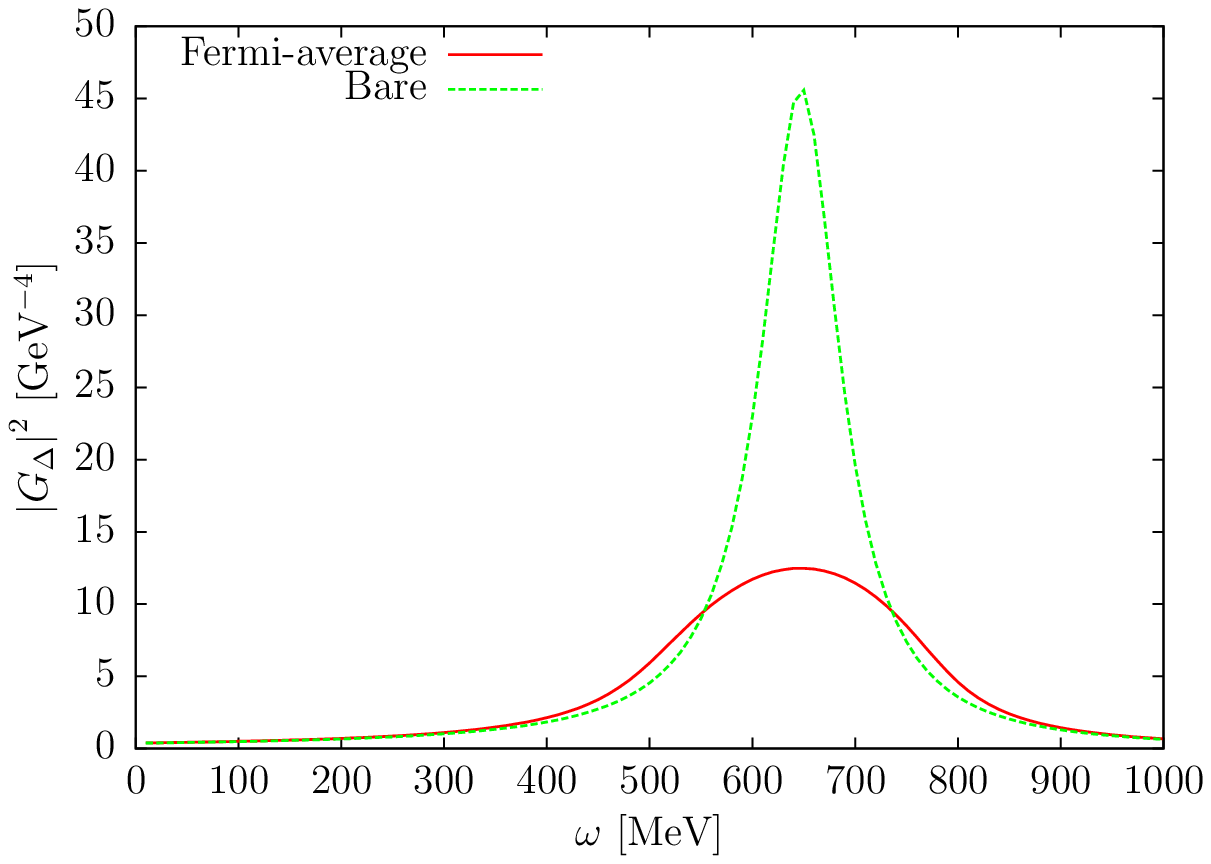}
\caption{(Color online) Square of the absolute value of the 
spin-independent term of the 
$\Delta$-propagator in frozen approximation compared to the 
average propagator. In this evaluation we have taken $q=1000$
MeV/c and $\Gamma_\Delta=120$ MeV.}\label{fig_denom}
\end{figure}

\section{Results}\label{sec_results}

In this section we validate the frozen approximation by computing the
approximate 2p-2h response functions and comparing with the exact
results in the RFG. We consider the case of the nucleus ${}^{12}$C
with Fermi momentum $k_F=225$ MeV/c, and show the different response
functions for low to high values of the momentum transfer.  For other
nuclei with different $k_F$ the frozen parameters of Table 1 should be
determined again, and we expect their values change slightly.

In Fig.~\ref{fig_denom} we show the modulus squared of the $\Delta$
propagator, given by the $G_\Delta(H+Q)$ function defined in
Eq.~(\ref{denominator}), computed for $\nh=0$, as a function of
$\omega$ for $q=1$ GeV/c.  It presents the typical Lorentzian shape
corresponding to width $\Gamma_\Delta=120$ MeV.  We observe a 
narrow peak around $\omega\simeq 650$ MeV. This corresponds to the
$\Delta$-peak position for $q=1$ GeV/c.  In the same figure we also show
the square of the frozen average $G_{\rm frozen}$ (solid line).
The resulting peak is quenched and broadened as compared to the
Lorentzian shape, reducing its strength and enlarging its width. This
behavior of the averaged $\Delta$-propagator drives the actual shape
of the exact 2p-2h nuclear responses, being more realistic than the
simple Lorentzian shape of the frozen approximation without the
average, as we will see below.

\begin{figure}[tbhp]
\centering
\includegraphics[width=7.5cm,bb=150 60 425 780]{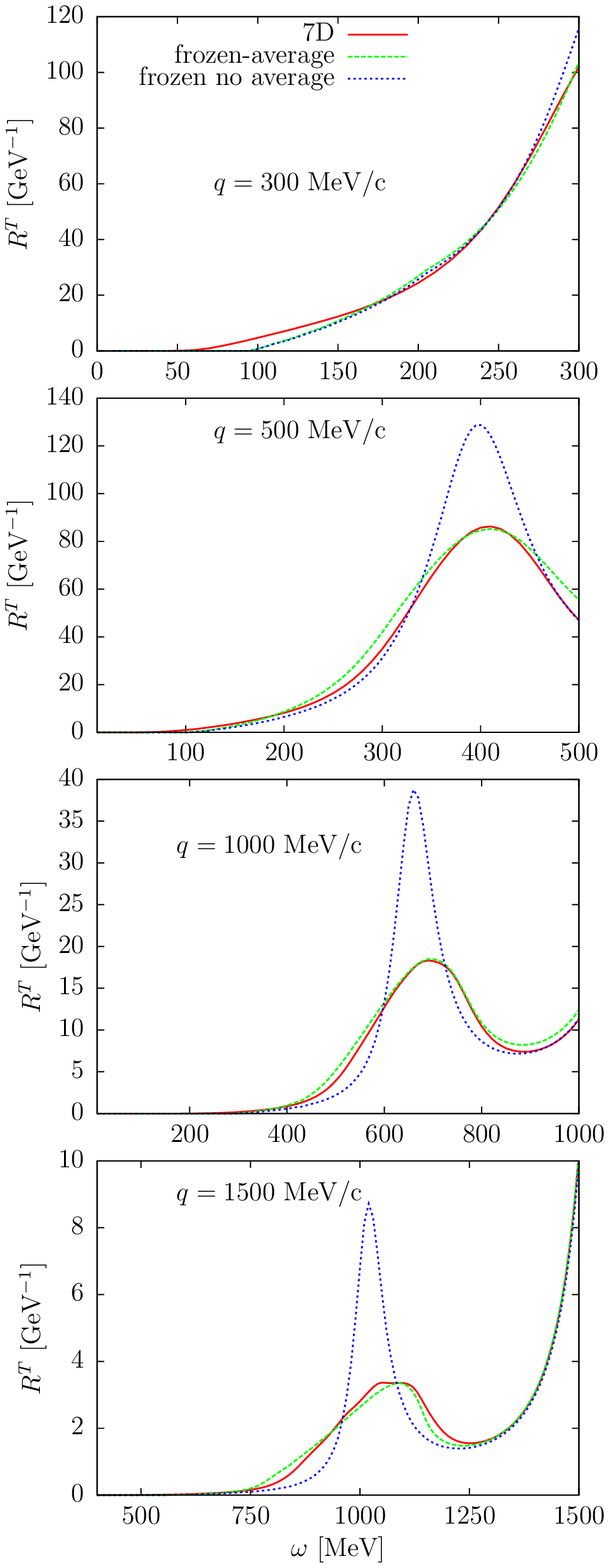}
\caption{(Color online) 2p-2h transverse response function $R^T$
of ${}^{12}$C for different momentum transfers $q$. 
The exact results are compared to the frozen approximation with and without the averaged $\Delta$ propagator.
}\label{fig_rt}
\end{figure}

In Fig.~\ref{fig_rt} we show the weak transverse 2p-2h 
response function of ${}^{12}$C for four different values of
the momentum transfer ranging from 300 to 1500 MeV/c. 
The curves correspond to different
calculations or approximations made in the evaluation of the
responses, as labeled in the legend. The solid line corresponds
to the seven-dimensional calculation with no approximations. 
The other two curves refer to the different frozen nucleon 
approximations developed in this work: 
the dashed line is obtained with Eq.~(\ref{frozen_eq}) but performing the replacement expressed
in (\ref{replacement_rule}) for the forward $\Delta$-excitation
terms in the evaluation of the current matrix elements;
on the contrary, the dotted line corresponds to the same 
frozen nucleon approximation, Eq.~(\ref{frozen_eq}),
but without the Fermi-average of the $\Delta$-propagator
in the forward terms.

As it can be seen from Fig.~\ref{fig_rt}, for those values
of the momentum transfer for which the $\Delta$-peak is not
reached (the panel with $q=300$ MeV/c), there is really little
difference between averaging or not the $\Delta$ propagator. 
This is certainly not the case when the $\Delta$-peak
is fully reached, as shown in the other panels. In this situation
there is a dramatic difference between performing the 
Fermi-average of the $\Delta$-propagator or not. This difference
is in consonance with the results shown in the previous Fig.~\ref{fig_denom}, 
and it shows how crucial is the treatment
of the $\Delta$-propagator to obtain accurate results for
the 2p-2h responses in the frozen nucleon approximation, i.e, 
with only one integration. 

The results in Fig.~3 have been obtained after fitting the
parameters $(\Delta_{\rm frozen},\, \Gamma_{\rm frozen})$ for the
Fermi-averaged $\Delta$-propagator at the different values of the
momentum transfer quoted in Table~\ref{table1}.  It is also worth
noting that there is no way of converting the dotted line into the
dashed one by only a suitable fitting of these parameters, {\it i.e.,}
without averaging the $\Delta$-propagator.

\begin{figure}[t]
\centering
\includegraphics[width=8cm,bb=180 425 430 780]{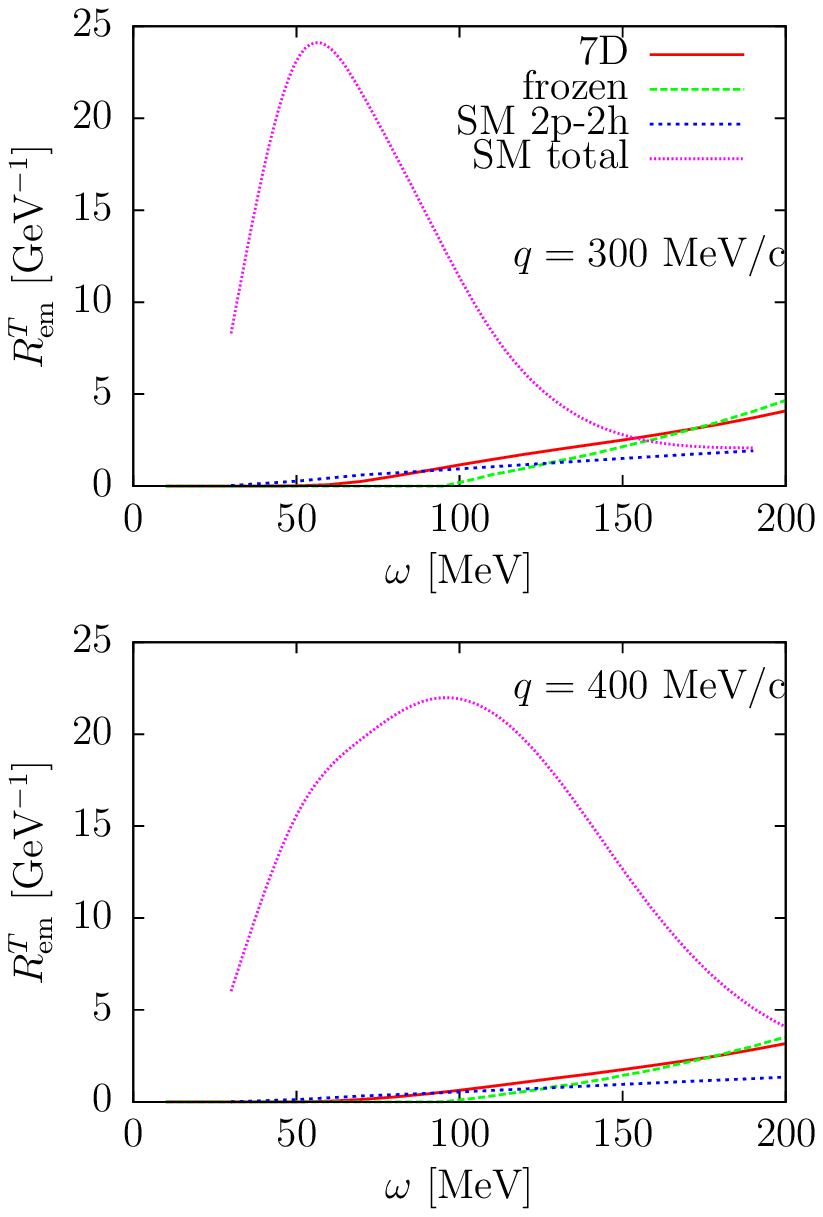}
\caption{(Color online) Comparison of 2p-2h electromagnetic transverse response
  functions of ${}^{12}$C within different models for two values of
  the momentum transfer. The exact RFG results and the frozen
  approximation are compared with the shell model (SM) results of
  \cite{Amaro:1994fx}.  The total shell model results (1p-1h) +
  (2p-2h) are also shown for comparison. 
}\label{fig_shellmodel}
\end{figure}

\begin{figure}[t]
\centering
\includegraphics[width=8cm, bb=150 220 450 780]{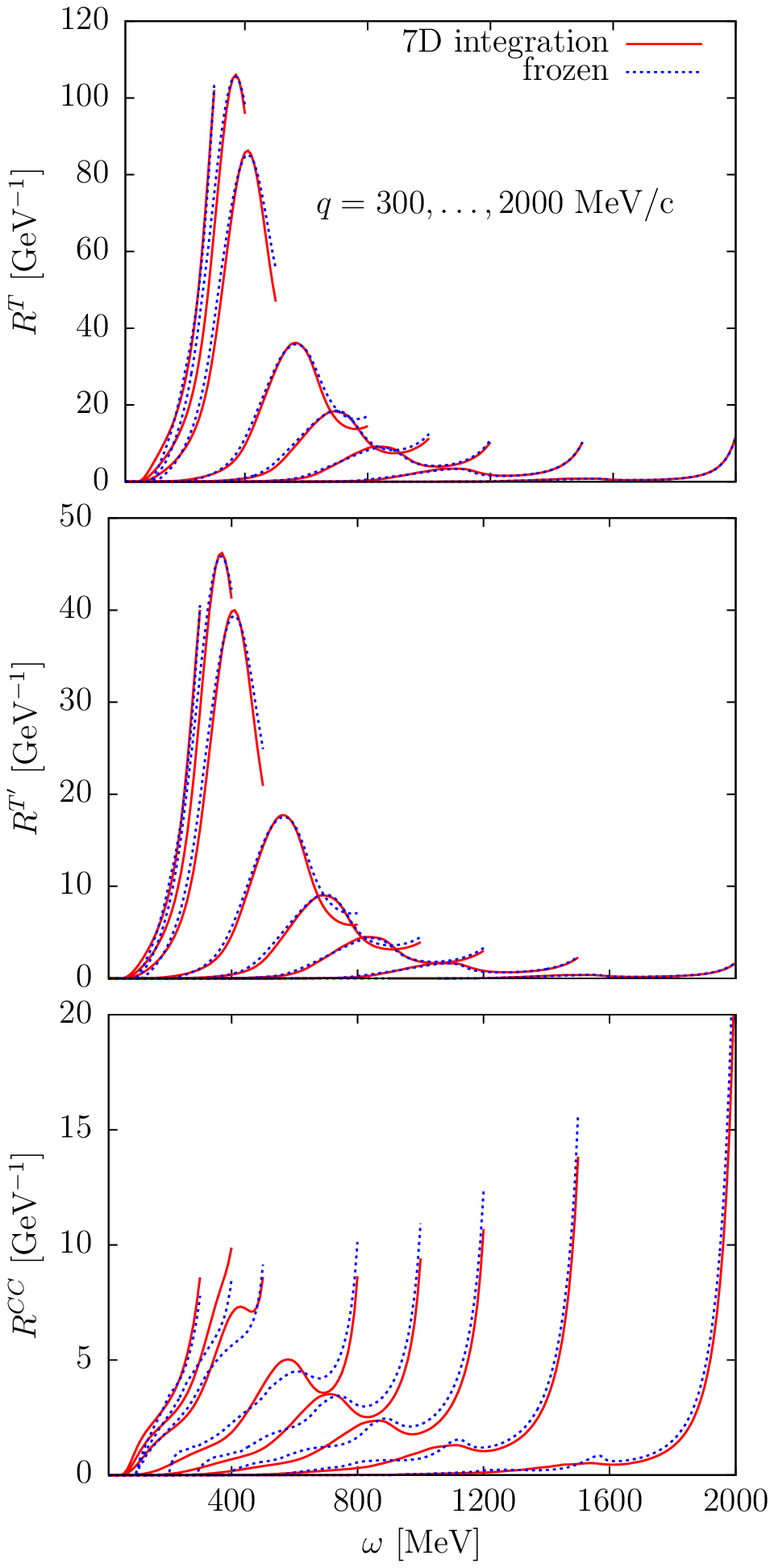}
\caption{(Color online) Comparison of the frozen approximation with the exact results
for several 2p-2h CC weak 
response functions. The target is ${}^{12}$C. Several 
values of the momentum transfer are displayed:
$q=300, 400, 500, 1000, 1200, 1500$ and 2000 MeV/c.
}
\label{fig5}
\end{figure}

In Fig.~\ref{fig_shellmodel} we show results for the transverse
electromagnetic 2p-2h response function. The frozen and exact (7D) $T$
response of the RFG are compared with the results obtained in the
shell model 2p-2h calculation of \cite{Amaro:1994fx}. 
This was one of the first computations of the 2p-2h response
within the nuclear shell model.  
The total nuclear response in the
shell model, obtained by adding the 1p-1h to the 2p-2h channel, is
also shown to appreciate the relative size of the 2p-2h contribution
to the total result.

As shown in Fig.~\ref{fig_shellmodel}, the Fermi gas
results (either in frozen approximation or not) are similar to the
shell model ones. The small discrepancy between them cannot be attributed to
relativistic effects because of the low momentum transfer values considered, but to the
different coupling constants and form factors used in the
model of the $\Delta$ meson-exchange current considered in
\cite{Amaro:1994fx} and the present approach.  We can remark the
slightly different threshold effects between both calculations. These
effects are, as expected, very sensitive to the treatment of the
nuclear ground state.  Note also that the frozen approximation
describes reasonably well this low momentum $q=300$ MeV/c, considering
the simplifications involved.

Finally in Fig.~\ref{fig5} we show that the frozen approximation works
notably well in a range of momentum transfer from low to high values
of $q$. We compare the $T$, $T'$, and $CC$ 2p-2h responses in frozen
approximation with the exact results obtained computing numerically
the 7D integral of the hadronic tensor. The accord is particularly
good for the two transverse responses which dominate the cross
section.  A slight disagreement occurs for very low energy transfer at
threshold where the response functions are anyway small. In the case
of the CC response function some tiny differences are observed.
However, note that this response is small because the dominant
$\Delta$ current is predominantly transverse. Moreover, its global
contribution to the cross section is not very significant because it
is partially canceled with the contribution of the CL and LL
responses.

\paragraph {Physical Interpretation of the frozen approximation}  The
validity of the frozen approximation led us to conclude that, in the
inclusive responses for two-particle emission, the detailed
information about the momenta carried out by the two nucleons
is lost.  This is because the energy and momentum transfer $q,\omega$
are shared by the two nucleons in multiple ways. This is reminiscent
from the phase-space kinematical dependence (which can be obtained
setting the elementary
response $r^K$ to unity) already seen in \cite{Simo:2014wka}. The
soft dependence of the elementary response on the initial momenta
makes the same argument applicable to the full responses with the
exception of the $\Delta$ forward current that requires one to soften and
average the rapid variations of the $\Delta$ propagator. Only the low-energy 
region where the sharing is highly restricted and the cross
section is therefore very small, is found to be sensitive to the details of
the initial state. This is also supported by the comparison between
the shell model and the RFG. 

On the other hand, in the 2p-2h model of \cite{Cuy16,Cuy17}, an average
momentum $\np'_2= \nq-\np'_1$ was determined by imposing
quasi-deuteron kinematics. Note that this condition is  similar to the
present frozen approach, but this only guarantees that the total
momentum of the two holes is zero, corresponding to selecting
back-to-back pair configurations in the ground state only.

\section{Conclusions}\label{sec_conclusions}

In this work we have introduced and validated the frozen nucleon
approximation for a fast and precise calculation of the inclusive
2p-2h response functions in a relativistic Fermi gas model.  This
approximation neglects the momentum dependence of the two holes in the
ground state and requires the use of an effective propagator for the
$\Delta$ resonance conveniently averaged over the Fermi sphere, for
which we have provided a simple analytical expression. For momentum
transfers above the Fermi momentum this approximation makes it
possible to compute the responses with only a one-dimensional
integral. Taking into account all the uncertainties in modeling the
two-nucleon emission reactions, this approach can be used instead of
the full 7D integral, obtaining very satisfactory results.  Although
we have used a specific model of MEC to prove the validity of the
approximation, it is reasonable to expect that the frozen approach is
also valid for other 2p-2h models. This can be of great interest when
implementing 2p-2h models in Monte Carlo event generators, which up to
now have relied on parameterizations from external calculations. In
summary, the frozen approximation enables one to make 2p-2h
calculations very efficiently and rapidly, instead of interpolating
pre-calculated tables, including allowing the parameters of the models
to be modified inside the codes, if desired. Finally, in the near
future this study will be extended to an exploration of how the 2p-2h
MEC responses depend on nuclear species \cite{kfpaper}.

\section{Acknowledgements}

This work has been partially supported by the Spanish Ministerio de
Economia y Competitividad and ERDF (European Regional Development
Fund) under contracts FIS2014-59386-P, FIS2014-53448-C2-1, by the Junta de
Andalucia (grants No. FQM-225, FQM160), by the INFN under project
MANYBODY, and part (TWD) by the U.S. Department of Energy under
cooperative agreement DE-FC02-94ER40818. IRS acknowledges support from
a Juan de la Cierva fellowship from MINECO (Spain). GDM acknowledges
support from a Junta de Andalucia fellowship (FQM7632, Proyectos de
Excelencia 2011).


\begin{thebibliography}{99}


\bibitem{Mos16} U. Mosel, Ann. Rev. Nuc. Part. Sci. 2016. 66:1-26.

\bibitem{Kat17}
  T.~Katori and M.~Martini,
  arXiv:1611.07770 [hep-ph].


\bibitem{Nie11} J. Nieves, I. Ruiz Simo, M.J. Vicente Vacas, 
Phys.Rev. C83 (2011) 045501.

\bibitem{Nie12} J. Nieves, I. Ruiz Simo, M.J. Vicente Vacas, 
Phys.Lett. B707 (2012) 72.

\bibitem{Gra13} R. Gran, J. Nieves, F. Sanchez, M.J. Vicente Vacas, 
Phys.Rev. D 88 (2013) 113007.

\bibitem{Ama11} 
J.E. Amaro, M.B. Barbaro, J.A. Caballero, T.W. Donnelly, 
C.F. Williamson,
    Phys.Lett. B696 (2011) 151.

\bibitem{Ama12} J.E. Amaro, M.B. Barbaro, J.A. Caballero, T.W. Donnelly, 
Phys.Rev.Lett. 108 (2012) 152501.

\bibitem{Mar09} M. Martini, M. Ericson, G. Chanfray, J. Marteau, 
                 Phys.Rev. C80 (2009) 065501.

\bibitem{Mar10} M. Martini, M. Ericson, G. Chanfray, J. Marteau, 
Phys.Rev. C81 (2010) 045502.

\bibitem{Agu10}
A.A. Aguilar-Arevalo, et al. (MiniBooNE Collaboration),  
Phys.Rev. D81 (2010) 092005. 

\bibitem{Nak11} SciBooNE Collaboration (Nakajima, Y. et al.),
 Phys.Rev. D83 (2011) 012005.

 \bibitem{And12} Anderson, C. et al. (ArgoNeuT Collaboration), 
 Phys.Rev.Lett. 108 (2012) 161802.

\bibitem{Abe13} K. Abe et al. (T2K Collaboration), 
                Phys.Rev. D87 (2013) 9, 092003. 
 
\bibitem{Fio13} G.A. Fiorentini et al. (MINERvA Collaboration), 
Phys.Rev.Lett. 111 (2013) 022502.

\bibitem{Abe14} Abe, K. et al. (T2K Collaboration), 
 Phys.Rev. D90 (2014) 5, 052010. 

\bibitem{Wal15} Walton, T. et al. (MINERvA Collaboration), 
 Phys.Rev. D91 (2015) 7, 071301.

\bibitem{Ank16} A.M. Ankowski, O. Benhar, C. Mariani, E. Vagnoni,
  arXiv:1603.01072.

\bibitem{Rod16} P.A. Rodrigues, et al. (MINERvA Collaboration), 
Phys. Rev. Lett. 116, 071802 (2016).

\bibitem{Hayato:2009zz}
  Y.~Hayato,
  Acta Phys.\ Polon.\ B {\bf 40} (2009) 2477.

\bibitem{Andreopoulos:2009zz}
  C.~Andreopoulos [GENIE Collaboration],
  Acta Phys.\ Polon.\ B {\bf 40} (2009) 2461.

\bibitem{Andreopoulos:2009rq}
  C.~Andreopoulos {\it et al.},
  Nucl.\ Instrum.\ Meth.\ A {\bf 614} (2010) 87.

  
\bibitem{Sobczyk:2008zz}
  J.~Sobczyk,
  PoS NUFACT {\bf 08} (2008) 141.

\bibitem{Nieves:2011pp}
  J.~Nieves, I.~Ruiz Simo and M.~J.~Vicente Vacas,
  Phys.\ Rev.\ C {\bf 83} (2011) 045501.


\bibitem{Gran:2013kda}
  R.~Gran, J.~Nieves, F.~Sanchez and M.~J.~Vicente Vacas,
  Phys.\ Rev.\ D {\bf 88} (2013) no.11,  113007.


\bibitem{Schwehr:2016pvn}
  J.~Schwehr, D.~Cherdack and R.~Gran,
  arXiv:1601.02038 [hep-ph].

\bibitem{Martini:2009uj}
  M.~Martini, M.~Ericson, G.~Chanfray and J.~Marteau,
  Phys.\ Rev.\ C {\bf 80} (2009) 065501.

\bibitem{Bodek:2011ps}
  A.~Bodek, H.~S.~Budd and M.~E.~Christy,
  Eur.\ Phys.\ J.\ C {\bf 71} (2011) 1726.

\bibitem{Zmuda:2015twa}
  J.~\.Zmuda, K.~M.~Graczyk, C.~Juszczak and J.~T.~Sobczyk,
  Acta Phys.\ Polon.\ B {\bf 46} (2015) no.11,  2329.


\bibitem{VanOrden:1980tg}
  J.~W.~Van Orden and T.~W.~Donnelly,
  Annals Phys.\  {\bf 131} (1981) 451.

\bibitem{Simo:2014wka}
  I.~Ruiz Simo, C.~Albertus, J.~E.~Amaro, M.~B.~Barbaro, J.~A.~Caballero and T.~W.~Donnelly,
  Phys.\ Rev.\ D {\bf 90} (2014) no.3,  033012.

\bibitem{Simo:2016ikv}
  I.~Ruiz Simo, J.~E.~Amaro, M.~B.~Barbaro, A.~De Pace, J.~A.~Caballero and T.~W.~Donnelly,
  arXiv:1604.08423 [nucl-th].

\bibitem{Amaro:2005dn}
J.~E.~Amaro, M.~B.~Barbaro, J.~A.~Caballero, T.~W.~Donnelly and C.~Maieron,
  Phys.\ Rev.\ C {\bf 71} (2005) 065501.


\bibitem{Simo:2014esa}
  I.~Ruiz Simo, C.~Albertus, J.~E.~Amaro, M.~B.~Barbaro, J.~A.~Caballero and T.~W.~Donnelly,
  Phys.\ Rev.\ D {\bf 90} (2014) no.5,  053010


\bibitem{Hernandez:2007qq}
  E.~Hernandez, J.~Nieves and M.~Valverde,
  Phys.\ Rev.\ D {\bf 76} (2007) 033005.

\bibitem{Amaro:1994fx}
  J.~E.~Amaro, A.~M.~Lallena and G.~Co,
  Nucl.\ Phys.\ A {\bf 578} (1994) 365.


\bibitem{Cuy16} 
  T.~Van Cuyck, N.~Jachowicz, R.~Gonzalez-Jimenez, M.~Martini, V.~Pandey, J.~Ryckebusch and N.~Van Dessel,
  Phys.\ Rev.\ C {\bf 94}, no. 2, 024611 (2016)


\bibitem{Cuy17}
  T.~Van Cuyck, N.~Jachowicz, R.~Gonzalez-Jimenez, J.~Ryckebusch and N.~Van Dessel,
  arXiv:1702.06402 [nucl-th].


  \bibitem{kfpaper}  In preparation.  
  
\end{thebibliography}
\end{document}